\documentclass{ws-ijmpd}

\begin{document}

%
%

\title{VERITAS: STATUS c. 2009}

\author{T.~C.~Weekes}

\address{Whipple Observatory, Harvard-Smithsonian Center for Astrophysics\\
P.O. Box 6369, Amado, Arizona 85629-0097,U.S.A.\\
tweekes@cfa.harvard.edu}

\author{V.~A.~Acciari,
T.~Arlen,
T.~Aune,
M.~Beilicke,
W.~Benbow,
D.~Boltuch,
S.~M.~Bradbury,
J.~H.~Buckley,
V.~Bugaev,
K.~Byrum,
A.~Cannon,
A.~Cesarini,
L.~Ciupik,
Y.~C.~Chow,
P.~Cogan,
W.~Cui,
C.~Duke,
S.~J.~Fegan,
J.~P.~Finley,
G.~Finnegan,
P.~Fortin,
L.~Fortson,
A.~Furniss,
N.~Galante,
D.~Gall,
G.~H.~Gillanders,
S.~Godambe,
J.~Grube,
R.~Guenette,
G.~Gyuk,
D.~Hanna,
J.~Holder,
D.~Horan,
C.~M.~Hui,
T.~B.~Humensky,
A.~Imran,
P.~Kaaret,
N.~Karlsson,
M.~Kertzman,
D.~Kieda,
A.~Konopelko,
H.~Krawczynski,
F.~Krennrich,
M.~J.~Lang,
G.~Maier,
S.~McArthur,
A.~McCann,
M.~McCutcheon,
J.~Millis,
P.~Moriarty,
R.~A.~Ong,
A.~N.~Otte,
D.~Pandel,
J.~S.~Perkins,
A.~Pichel,
M.~Pohl,
J.~Quinn,
K.~Ragan,
L.~C.~Reyes,
P.~T.~Reynolds,
E.~Roache,
H.~J.~Rose,
M.~Schroedter,
G.~H.~Sembroski,
A.~W.~Smith,
D.~Steele,
S.~P.~Swordy,
M.~Theiling,
S.~Thibadeau,
J.~A.~Toner,
A.~Varlotta,
V.~V.~Vassiliev,
S.~Vincent,
R.~G.~Wagner,
S.~P.~Wakely,
J.~E.~Ward,
A.~Weinstein,
D.~A.~Williams,
S.~Wissel,
M.~Wood,
and B.~Zitzer}

\address{VERITAS Collaboration (see 
http://veritas.sao.arizona.edu/conferences/authors/authors?icrc 
for list of institutions)}

\maketitle

\begin{history}
\received{18 December 2009}
\end{history}

\begin{abstract}
VERITAS is a ground-based gamma-ray observatory that
uses the imaging atmospheric Cherenkov technique and operates in the
very high-energy (VHE) region of the gamma-ray spectrum from 100 GeV to 50 TeV.
The observatory consists of an array of four 12m-diameter imaging 
atmospheric Cherenkov telescopes located in southern Arizona, USA.
The four-telescope array has been fully operational since 
September 2007, and over
the last two years, VERITAS has been operating with high 
reliability and sensitivity.
It is currently one of the most sensitive VHE observatories.
This paper summarizes the status of VERITAS 
as of October, 2009, and describes the detection
of several new VHE gamma-ray sources.

\end{abstract}


\section{Overview}	

The presence of very high energy particles in the relativistic outflows
from gamma-ray bursters, from microquasars and from active galactic 
nuclei (AGN) can only be unambiguously established by the observation of
gamma rays or neutrinos. Unfortunately neutrino astronomy still
lacks the required sensitivity and satellite-based gamma-ray telescopes lose 
sensitivity above 100 GeV. Hence the best channel for the investigation
of the TeV content of relativistic outflows come from the use of 
ground-based gamma-ray telescopes. Here the imaging atmospheric
Cherenkov technique \cite{weekes2008} is particularly useful. The technique
was originally developed to probe the origin of the cosmic radiation
which is generally assumed to be tied to shock acceleration in
supernova remnants.

The observation of TeV gamma rays from AGN has been particularly
successful and emission from more than two dozen blazars is now 
well established. Recently emission has also been detected from
radio and starburst galaxies but it is the highly variable and
beamed emission from blazars that is of most interest to this
community.

There is also some evidence for TeV emission from microquasars
although the sample is small \cite{hessmicroq} \cite{magicmicroq}
\cite{veritasmicroq}. The evidence is strong for the
presence of high energy particle acceleration in the relativistic 
outflows. From the small sample observed it is clear that 
the Galactic analog to blazars (microblazars) is rare and not
a dominant feature of the Galactic relativistic sky. 

To date, there has been no definite detection of TeV gamma rays 
in gamma-ray bursts. However at cosmological distances the emitted
flux of TeV gamma rays would be strongly attenuated by photon-photon
pair production in extragalactic space. Also, if the emission only 
occurs during the few seconds of prompt gamma-ray activity, then it
is only with serendipity that an imaging telescope would have the
burst within its limited field of view. All-sky ground-based telescopes 
have relatively poor sensitivity for the detection of such transients.

\section{Operational Status of VERITAS}

The
VERITAS (Very Energetic Radiation Imaging Telescope Array
System) observatory, located at the basecamp of the
Fred Lawrence Whipple Observatory in southern Arizona, USA,
was completed in June 2007.  
 
\begin{figure}[pt]
\centerline{\psfig{file=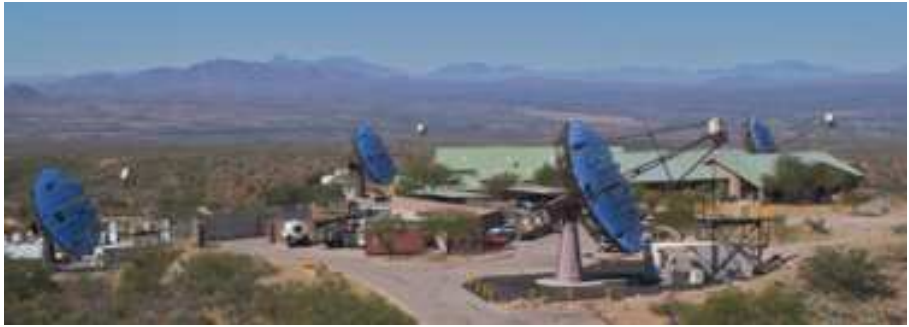,width=12.0cm}}
\vspace*{8pt}
\caption{The four 12m aperture reflectors that constitute the VERITAS
Observatory, shown after the relocation of one telescope in the summer
of 2009. The reflectors are conveniently distributed around the 
Administration Building of the Smithsonian's Whipple Observatory
in southern Arizona. \label{f1}}
\end{figure}

The observatory (Figure~\ref{f1}) consists of four 12m-diameter imaging 
atmospheric Cherenkov telescopes, with a typical baseline between telescopes
of $\sim$100\,m \cite{weekes02}.  
Each telescope has a
499-photomultiplier tube (PMT) camera, spanning
a field of view of $3.5^\circ$.
The signal from each camera pixel is amplified and recorded by a 
separate 500 MS/s Flash-ADC channel.
VERITAS employs a three-level trigger system;
Level 1 corresponds to the discriminators on each pixel, 
Level 2 is a pattern trigger for each telescope,
and Level 3 is the array trigger 
\cite{VERITAS}.

Regular observations with the full four-telescope array
started in September 2007, with
approximately 1000 h per year of observations taken.
The array has operated extremely well during the
last two years; more than 95\% of the observations have all
four telescopes operational.
The ability to take scientifically useful data under
partial moonlight was an important development -- it
adds approximately 30\% to the annual data yield.
At the recent ICRC a comprehensive review of the status
of VERITAS was given by R. Ong \cite{ICRC_ong}; this brief 
report is an abbreviated summary of the same material but 
is slightly updated.

VERITAS has an angular resolution
(68\% containment) of $< 0.1^\circ$,
a pointing accuracy of $< 50\,$arc-secs,
an energy range of 100\,GeV--50\,TeV, and
an energy resolution (above 200\,GeV) of 15--20\%.

\begin{figure}[pt]
\centerline{\psfig{file=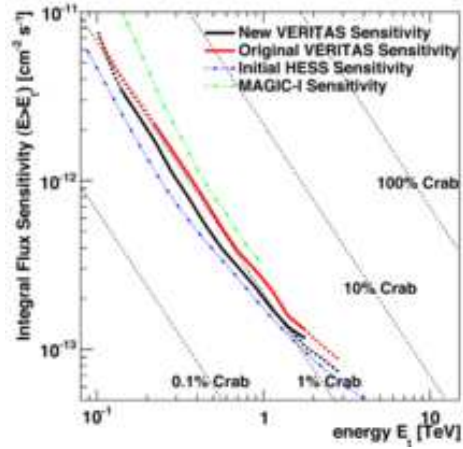,width=6.7cm}}
\vspace*{8pt}
\caption{Measured Integral Sensitivity of VERITAS as of
October, 2009 \label{fx} showing the effect of the recent upgrade; also
shown is that of HESS when built (with newly 
coated mirrors) and MAGIC.}
\end{figure}

In the summer of 2009, two important upgrades were made to VERITAS.
Firstly, Telescope 1 was moved to give a better array configuration.
Secondly, the facet alignment on individual telescopes was refined
\cite{hanna} to give better image definition. With the
combined effect of these upgrades,
the integral flux sensitivity is as shown in Figure~\ref{fx}.
The measured gamma-ray point source sensitivity of VERITAS
in its new configuration corresponds to 
the detection of a 1\% Crab Nebula source
at the five standard deviation level
in less than 25 h.

\section{Extragalactic Sources}

\subsection{M82}

A significant addition to the VHE zoo was the detection of the 
prototypical starburst galaxy, M82. 
Although the conventional picture of cosmic ray origins is diffusive shock
acceleration in isolated supernova remnants (SNR) \cite{hillas}, it has been
argued \cite{butt} that acceleration in superbubbles, regions of high
cosmic ray density, formed
by multiple supernovae explosions, is another viable source.
There is already some evidence for such emission
from star-forming regions of the Galaxy. 
The active regions of starburst galaxies (which contain an un
usually large number
of SNR) are also likely sources of cosmic ray acceleration
and are prime targets for the possible
emission of VHE gamma rays. 
The large size and activity level
of starburst galaxies compensates for their distance and permits a
calorimetric measurement of the expected high cosmic ray density.
 Because of its relative proximity the brightest
such starburst galaxy is expected to be M82; it is considered the prototype
of such galaxies.

\begin{figure}[pb]
\centerline{\psfig{file=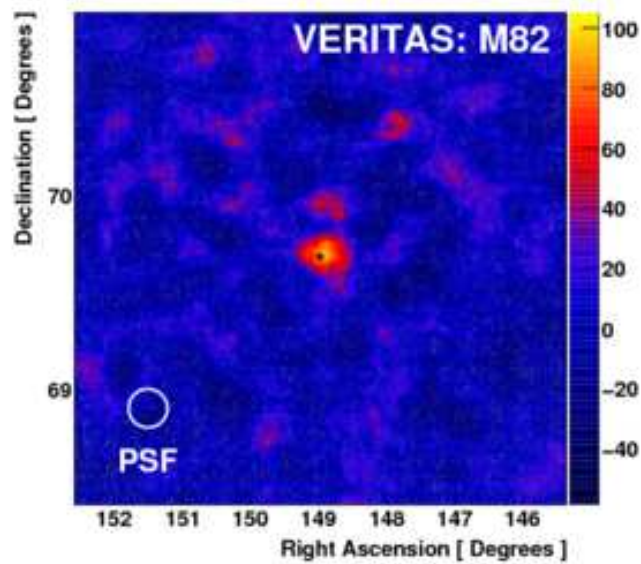,width=8.7cm}}
\vspace*{8pt}
\caption{Sky map in TeV gamma rays as seen by VERITAS in the vicinity
of M82. The star symbol represents the position of the galaxy. The PSF
(69\%) has the dimension shown in bottom left corner; clearly the
source is not resolved. \label{f9}}
\end{figure}

M82 was observed with VERITAS for a total of 137 hours of good quality
data between January, 2008 and April, 2009. Using the standard analysis
for low flux, hard spectrum, sources a signal at the 4.8 sigma level
(Figure~\ref{f9}) was detected.
The flux level is 0.9\% that of the Crab Nebula and, as such, is the
weakest source detected by VERITAS thus far.
The observed differential gamma-ray flux
is best fitted with a power law with index 
$\Gamma = 2.5 \pm 0.6_{\rm stat} \pm 0.2_{\rm sys}$
The gamma-ray luminosity above 700 GeV is 2 x 10$^{32}$ W or 2 x 10$^6$
times smaller than the infrared luminosity \cite{acciarinature}.

There is a correlation between the far infrared emission (thermal radiation
from warm dust) and radio emission (from synchrotron radiation from
cosmic-ray electrons) in starburst galaxies. The gamma-ray flux from M82
provides an estimate of the cosmic ray density which is directly associated
with both radiation mechanisms.

\subsection{M87}

A summary of observations by VERITAS and other VHE observatories is
given elsewhere by M. Belicke in these proceedings. It is of interest
that the three objects: M1, M82 and M87, that were the prime targets 
of observations with the atmospheric Cherenkov technique some four
decades ago \cite{fazio} have now all been detected.

\subsection{Blazars}

According to the canonical model, gamma rays are produced in 
relativistic jets which are powered by accretion unto the supermassive 
black holes in the centers of AGN. When the jet is pointed in the direction
of the observer it is defined as a blazar. The spectral energy distribution
(SED) of these objects is characterized by a double-humped structure, the 
first (low energy) peak is due to synchrotron radiation from high energy
electrons in the jet, the second (high energy) peak more likely from the
inverse-Compton scattering of these same electrons. If the low energy 
peak is at optical-infrared frequencies, then the source is designated a
low frequency BL Lac (LBL) and if, at X-rays frequencies, it is called
a high frequency BL Lac (HBL). Generally the bulk of the AGN detected at
MeV-GeV energies are LBLs and those at GeV-TeV energies are HBLs. There
is however a continuous distribution of properties and VERITAS has 
recently detected several intermediate frequency BL Lacs (IBLs).

\subsubsection{W Comae}

\begin{figure}[pb]
\centerline{\psfig{file=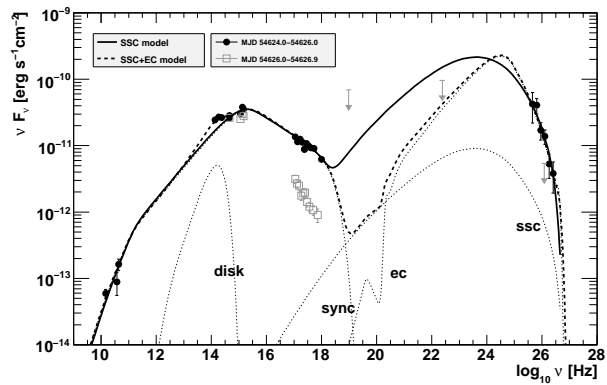,width=8.7cm}}
\vspace*{8pt}
\caption{SED of W Comae during its high state. Data comes from
VERITAS, Swift and radio and optical observatories. Synchrotron self
Compton and external synchrotron model fits are shown.  \label{f2}}
\end{figure}

This was the first IBL detected and appears to be highly variable.
It was first apparent as a flare over four nights in March, 2008 
(9\% Crab at peak) but is also seen as a weak steady source \cite{wcomae1}.
 It flared
again three months later \cite{wcomae2}. The spectrum is very soft with 
index 3.8$\pm$0.4.
The SED is shown in Figure~\ref{f2}.

\subsubsection{3C66a}

\begin{figure}[pt]
\centerline{\psfig{file=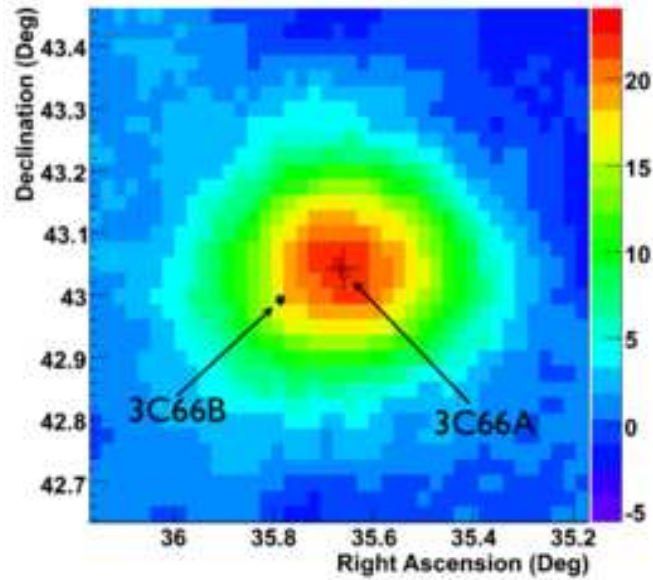,width=8.7cm}}
\vspace*{8pt}
\caption{Skymap of TeV gamma rays as seen by VERITAS in the
vicinity of 3C66a. Note that the association with 3C66a is clearly
favored over that for 3C66b. \label{f4}}
\end{figure}

VERITAS detected 3C66a when it was flaring in 2008 \cite{3C66a}. 
It is the second IBL detected by VERITAS. 
An earlier report by the Crimean group in 2002 \cite{crimea}
suggested it was at a very high level of emission. A signal from this region
was also reported by the MAGIC group \cite{MAGIC} but they suggested an
identification of the source with 3C66b, a nearby radio galaxy. The VERITAS
sky map (Figure~\ref{f4}) shows the peak of the TeV emission is very close 
to the nominal position of 3C66a (but
not consistent with 3C66b at the 4.8 sigma
level). The spectral index is very soft and 
is consistent with absorption
by extragalactic photons if the distance is that indicated by the published
redshift of 0.444. However, as with most BL Lac, there is some uncertainty 
about this value.

\subsubsection{PKS 1424+240}

PKS 1424+240 is a BL Lac object of unknown redshift. It was detected 
by VERITAS in 
the spring of 2009 (ATEL \#2084). No variations were seen. Similarly 
Fermi showed no variability was detected. The VERITAS signal is 
about  5\% of the Crab and it detected at the
$\sim$7.5 sigma level.The spectral index of the
VERITAS measurements above 140 GeV is
$\Gamma = 3.8 \pm 0.5_{\rm stat} \pm 0.3_{\rm sys}$.
The SED is well described by a one-zone synchrotron 
self-Compton model \cite{pks1424}.


\subsubsection{VER J0521+211}

VERITAS detected this source in October, 2009 (ATEL \#2260). 
The observations were motivated
by the detection of a $>$ 30 GeV source by Fermi at this position which is
consistent, within errors, with the radio-loud active galaxy RGB J0521.8+2112
of unknown redshift. At discovery the flux was $\sim$5\% of the Crab flux but
rose to 3-4 times this value the following month (ATEL \#2309).

\subsubsection{1ES 0502+675}

The detection of TeV gamma rays from the HBL BL Lac object, 1ES 0502+675,
is particularly important because of the apparent large red-shift (z = 0.341)
of the object (ATEL \#2301). It was seen by VERITAS in 13 hours of data
taken in September-November, 2009. The observations were motivated by the
flux and spectrum reported by Fermi. The VERITAS flux was about 4\% of the
Crab.   

\section{Galactic Sources}

Here we report on new detections by VERITAS of
several Galactic sources. A summary of VERITAS observations of microquasars
is reported by A. Smith elsewhere in these proceedings.

\subsection{G54.1+0.3}

The supernova remnant, G54.1+0.3,
was observed by VERITAS for 31\,h in 2008,
yielding a solid detection at the 6.8$\sigma$ level.
The VHE emission is consistent with a point source
at the pulsar location and there is no evidence for variability.
The flux level is $\sim$2.5\% Crab Nebula
above 1 TeV.
The differential spectral index is
$\Gamma = 2.4 \pm 0.2_{\rm stat} \pm 0.3_{\rm sys}$
\cite{G54}. The source is most likely associated with the
Pulsar Wind Nebula in G54.1+0.3.

\subsection{G106.3+2.7 (Boomerang)}

The supernova remnant G106.3+2.7 is 
part of a complex system that may have been
created by a supernova explosion occurring 
in a previously existing HI bubble.
The energetic pulsar associated with this
system, PSR J2229+6114,
has an estimated age of $\sim$10,000 years and
a spin-down luminosity of 
$\dot{\it E} \sim 2.2 \times 10^{37}$\,erg/s.
The SNR is within the error box of the
EGRET source 3EG J2227+6112, and the pulsar
appears on the Fermi Bright Source List
\cite{Cygnus_Fermi}.
Milagro reported $>$10 TeV emission from the general
region with a large error
box $\sim 1^\circ$ in diameter.

\begin{figure}[pt]
\centerline{\psfig{file=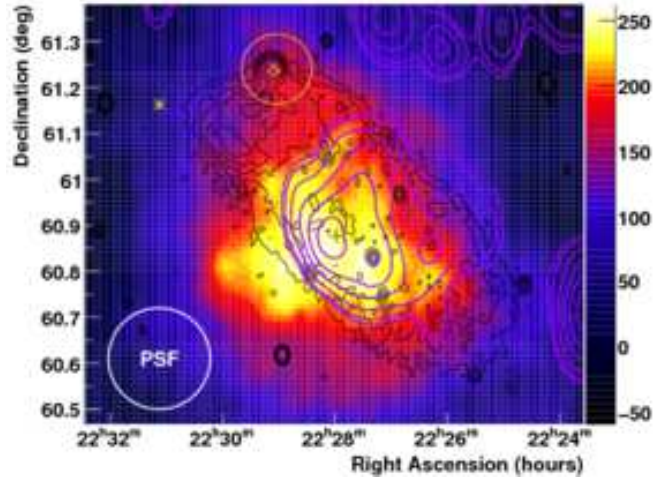,width=8.7cm}}
\vspace*{8pt}
\caption{Sky map of TeV gamma rays from G106.3+2.7 as seen by
VERITAS.The circle at top indicates the error contour for
the source seen by Fermi. The radio and CO contours are shown as
continuous lines.
The peak of the gamma-ray emission is
displaced from the pulsar and instead overlaps with
a region of high CO density. 
The measured VHE spectrum,
with differential spectral index
$\Gamma = 2.3 \pm 0.3_{\rm stat}
\pm 0.3_{\rm sys}$, is relatively hard and
is consistent with a power-law form up to
the Milagro energy of 35 TeV.
The spectrum and the observed morphology of the
source support a possible hadronic origin
for the VHE emission.
 \label{f7}}
\end{figure}

The VERITAS detection of VHE emission came
from 33 h of observations in 2008 
resulting in a post-trials significance of
6.0 sigma and an integral gamma-ray flux
level of $\sim$5\% Crab Nebula above 
1 TeV \cite{Boomerang_VERITAS}.
As shown in Figure~\ref{f7}, the
VHE emission is clearly extended spanning
a region approximately $0.4^\circ$ by $0.6^\circ$ in size.

\subsection{Cassiopeia A}

Cassiopeia A (Cas-A) is now a
well established VHE gamma-ray source.
Observations of Cas-A by VERITAS in 2007 yielded a clear detection at the
8.3$\sigma$ statistical level \cite{Humensky2}.
The integral gamma-ray flux is
$\sim$3.5\% of the Crab Nebula above 1 TeV.
The VERITAS energy spectrum is well fit by a power-law
with differential spectral index
$\Gamma = 2.6 \pm 0.3_{\rm stat}
\pm 0.2_{\rm sys}$ and there is no
indication of a cut-off at high energy.
There is also no evidence for any source
extension \cite{CAS_A_VERITAS}.

\subsection{IC443}

The emission of VHE gamma rays from IC 443
was first reported by MAGIC and VERITAS
in April 2007 at the VERITAS First Light Celebration.
MAGIC reported
a 5.7$\sigma$ detection of the source, corresponding
to an integral flux of $\sim$2.8\% Crab Nebula
above 300 GeV \cite{IC443_MAGIC}.
The VERITAS observations yield a 
statistical significance of 7.5$\sigma$ 
\cite{IC443_VERITAS} and
an integral flux of $\sim$3.2\% Crab Nebula,
consistent with the MAGIC result.

\begin{figure}[pt]
\centerline{\psfig{file=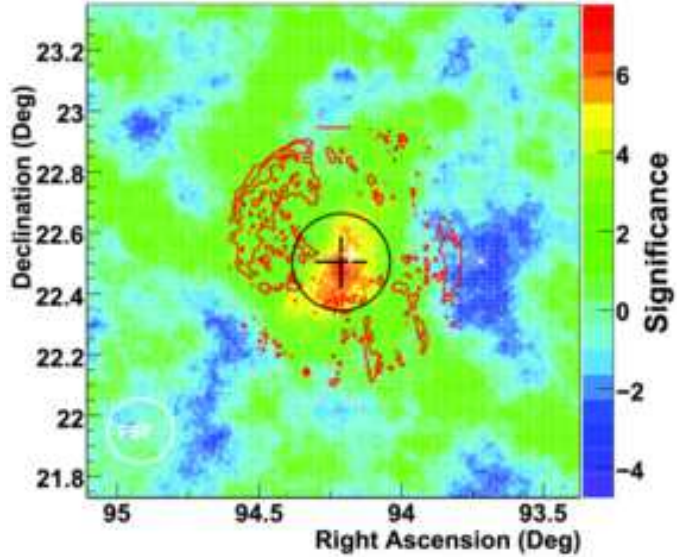,width=9.0cm}}
\vspace*{8pt}
\caption{Sky map of gamma rays as seen by VERITAS in the vicinity
of IC443.
The VERITAS data show that the gamma-ray emission is extended,
with a characteristic fitted
two-dimensional Gaussian radius of
$0.16^\circ$.
The VHE emission also overlaps with a dense
CO molecular cloud whose contours are shown as continuous lines. 
\label{f6}}
\end{figure}

\section{Outlook}

The four-telescope VERITAS array 
is operating extremely well ($> 95$\% uptime)
and with excellent sensitivity. 
VERITAS has detected 27 VHE gamma-ray
sources, ten previously not seen by
other instruments.
An upcoming upgrade program will further improve
the performance of VERITAS, ensuring that it
remain a premier gamma-ray observatory well
into the next decade.
 The upgrade is aimed at further improving the sensitivity and
extending the reach of VERITAS to lower energies.
The existing PMTs in each VERITAS camera will be replaced 
with ones having higher quantum efficiency.
A new topological telescope trigger system
is also envisioned.

\section{Acknowledgments}

This research is supported by grants from  
the U.S. National Science Foundation, 
the U.S. Department of Energy,
and the Smithsonian Institution, by NSERC in
Canada, by Science Foundation Ireland, and by STFC in the U.K.
The excellent work of the technical support staff at the FLWO and the
collaborating institutions is acknowledged.



\end{document}